\documentclass[aps,twocolumn,prb,groupedaddress]{revtex4}
\usepackage{color,graphicx}
\usepackage{amsmath}
\usepackage{natbib}
\usepackage{longtable}
\usepackage{gensymb}
\usepackage{pifont}
\usepackage{siunitx}

\begin{document}

\title{Patterning enhanced tetragonality in BiFeO$_3$ thin films with effective negative pressure by helium implantation}

\author{C. Toulouse$^{1,2}$}
\altaffiliation[\textbf{Present address :}]{ Physics and Materials Science Research Unit, University of Luxembourg, LIST-Belvaux site, 41 rue du Brill, L-4422 Belvaux, Luxembourg}
\altaffiliation[\\\textbf{Corresponding Author :}]{ constance.toulouse@uni.lu}
\author{J.~Fischer$^3$}
\author{S.~Farokhipoor$^4$}
%\author{S. Glin\v{s}ek$^1$}
%\author{P. Grysan$^1$}
\author{L.~Yedra$^5$}
\author{F.~Carl\`a$^{6,7}$}
\author{A.~Jarnac$^8$}
\author{E.~Elkaim$^8$}
\author{P.~Fertey$^8$}
%\author{J. I{\~n}iguez$^1$}
\author{J.-N.~Audinot$^2$}
\author{T.~Wirtz$^2$}
\author{B.~Noheda$^4$}
\author{V.~Garcia$^3$}
\author{S.~Fusil$^3$}
\author{I.~Peral~Alonso$^1$}
\author{M.~Guennou$^1$}
\author{J.~Kreisel$^1$}
%\affiliation{$^1$Materials for Research and Technology - Ferroic Materials for Transducers, Luxembourg Institute of Science and Technology, 41 rue du Brill, L-4422 Belvaux, Luxembourg\\
%$^2$Physics and Materials Science Research Unit - University of Luxembourg, 2 avenue de l'Université L-4365 Esch-sur-Alzette, Luxembourg\\
%$^3$Nanostructures of Functional Oxides - University of Groningen, Nijenborgh 4, 9747 AG, Groningen, The Netherlands\\
%$^4$European Synchrotron Radiation Facility, F-38043 Grenoble, France\\
%$^5$Unité Mixte de Recherche, 1 Avenue Augustin Fresnel, 91767 Palaiseau, France CNRS-Thales}

\affiliation{$^1$Department of Physics and Materials Science, University of Luxembourg, 41 rue du Brill, L-4422 Belvaux, Luxembourg\\
$^2$Materials Research and Technology Department, Luxembourg Institute of Science and Technology, 41 rue du Brill, L-4422 Belvaux, Luxembourg\\
$^3$Unité Mixte de Physique, CNRS, Thales, Université Paris-Saclay, 91767 Palaiseau, France\\
$^4$Zernike Institute for Advanced Materials, University of Groningen, Nijenborgh 4, 9747AG Groningen, The Netherlands\\
$^5$Laboratoire Structures, Propriétés et Modélisation des Solides (UMR CNRS 8580) and Laboratoire Mécanique des Sols, Structures et Matériaux (UMR CNRS 8579), CentraleSupélec, Univ. Paris Saclay, Gif-sur-Yvette, France\\
$^6$European Synchrotron Radiation Facility, Grenoble, France\\
$^7$Diamond Light Source Ltd, Didcot, United Kingdom\\
$^8$Synchrotron SOLEIL, L'Orme des merisiers, Saint-Aubin, Gif-sur-Yvette, France}

%\date{\today}

\begin{abstract}
Helium implantation in epitaxial thin films is a way to control the out-of-plane deformation independently from the in-plane strain controlled by epitaxy. In particular, implantation by means of a helium microscope allows for local implantation and patterning down to the nanometer resolution, which are of interest for device applications. We present here a study of bismuth ferrite (BiFeO$_3$) films where strain was patterned locally by helium implantation. Our combined Raman, XRD and TEM study shows that the implantation causes an elongation of the BiFeO$_3$ unit cell and ultimately a transition towards the so-called super-tetragonal polymorph via states with mixed phases. In addition, TEM reveals the onset of amorphization at a threshold dose that does not seem to impede the overall increase in tetragonality. The phase transition from the R-like to T-like BiFeO$_3$ appears as first-order in character, with regions of phase coexistence and abrupt changes in lattice parameters.  
\end{abstract}

\maketitle

\section{Introduction}
%Pertsev2000, Tyunina2010, Kim2011, Guo2018
%Negative pressure / helium implantation / Patterning power of HIM / Reversibility

Strain engineering has arisen in the last decade as an essential means to tune physical properties in functional thin films. The most common way to tune strain in films is by using epitaxial strain, imposing an in-plane biaxial strain to the material due to the lattice misfit with the substrate on which the film is grown. The strain state can be varied by using different substrates, or even in some instances, can be tuned continuously, for example by an electric field applied on a piezoelectric substrate\cite{bilani-zeneli_srtio3_2008, dorr_model_2009, herklotz_electrical_2010}. Using this approach, it has been demonstrated that the properties of the films can be controlled and modified, sometimes spectacularly. Remarkable results have included induced ferroelectricity\cite{pertsev_phase_2000, tyunina_evidence_2010, kim_epitaxial_2011, guo_strain-induced_2018}, a modified magnetic ground state of multiferroic compounds\cite{lee_epitaxial-strain-induced_2010, sando_crafting_2013, agbelele_strain_2016} or strain-induced structural transitions\cite{yamada_antiferrodistortive_2010, schwarzkopf_strain-induced_2012, mao_structural_2016, weber_multiple_2016}. 
In this classical approach of strain engineering, only the in-plane biaxial strain is controlled; the out-of-plane strain is fixed by the elastic equilibrium of the system, but cannot be controlled independently. 

Controlling this additional strain parameter, and in particular achieving an elongation of the unit cell (``negative pressure"), is of high interest. Negative pressure has, indeed, been theoretically predicted to trigger various properties modifications \cite{tinte_anomalous_2003, liu_negative_2009, wang_negative-pressure-induced_2015, kvasov_piezoelectric_2016}. Experimentally, it can be achieved by helium implantation. Due to its nobility, helium implants interstitially, without chemical substitution, thus inducing a `swelling' of the host material's unit cell volume\cite{sharma_designing_2017}.
Helium implantation is customarily achieved by lab ion sources \cite{guo_strain_2015, herklotz_continuously_2016, herklotz_controlling_2016, guo_strain-induced_2018, herklotz_designing_2019} similar to the ones used for wafer processing in semiconductor engineering. Another route - the one chosen in the present work - is to use a helium ion microscope as a way to implant ions locally.  
This technique has recently started to be used as a means of defect engineering\cite{mcgilly_nanoscale_2017, saremi_local_2018}, but our study is aiming at using a helium microscope for strain-engineering purposes.
The interest of this new method lies in particular in the nano-patterning possibilities allowed by the sub-nanometer resolution of the microscope\cite{zeiss_microscopy_2008}.
Moreover, it has been demonstrated recently that helium implantation is a reversible process and that the helium trapped can be released with annealing at high temperatures above \SI{650}{\celsius} and under oxygen atmosphere\cite{guo_strain_2015}, allowing for reversible properties tuning. For instance, a very recent study on FeRh thin films shows that local helium implantation allows for direct writing of nanoscale domains with a metamagnetic order tunable with the implanted dose, demonstrating the application potential of this technique\cite{cress_direct-write_2020}.

%BFO / Multiferroic ppties of bulk / structural ppties of films / biblio

%--------------------------------------------------------------------------------------------------
\begin{figure*}[t!]
\resizebox{\textwidth}{!}{\includegraphics{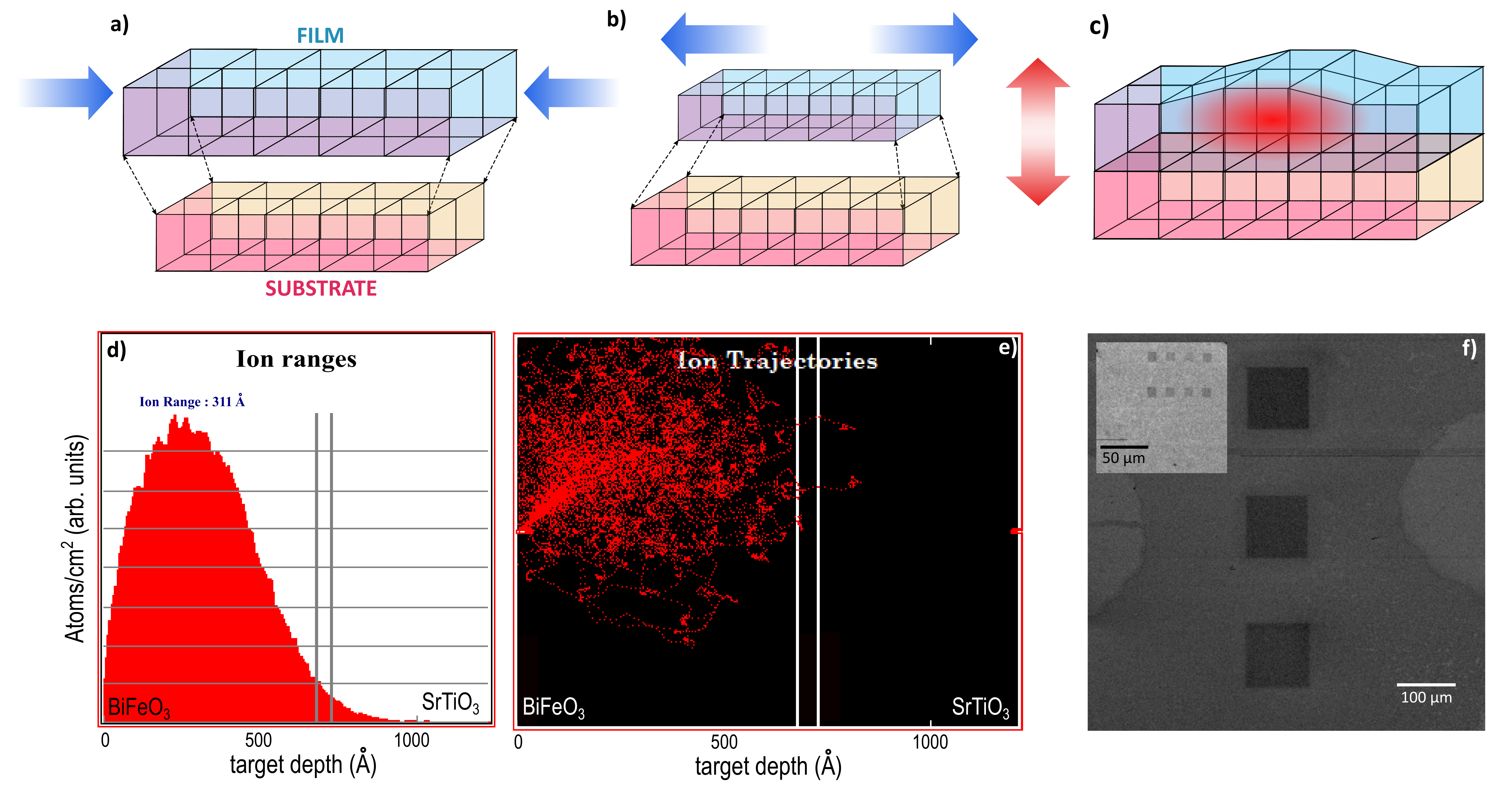}}
\caption{Schematic representations of (a) compressive and (b) tensile epitaxial strain and of (c) the out-of-plane negative pressure induced by helium implantation. TRIM simulations (see "Experimental details" section for details) of (d) the helium ion distribution and (e) the implantation depth profile of 6~keV helium ions inside BiFeO$_3$(\SI{70}{\nm})/SrRuO$_3$(\SI{5}{\nm})//SrTiO$_3$ with an incident angle of 49$\degree$. (f) Secondary electron images (SEM-HIM) of patterned BiFeO$_3$ films with various sizes of implanted regions.}
\label{NP-SRIM}
\end{figure*}
%--------------------------------------------------------------------------------------------------

Here, we demonstrate the use of a helium ion microscope to perform strain engineering by helium implantation on bismuth ferrite (BiFeO$_3$) thin films.
BiFeO$_3$ is one of the most studied multiferroic compounds as it exhibits multiferroicity well above room temperature. In its bulk form it develops below $T_\mathrm{C} = \SI{1143}{\K}$ a very large ferroelectric polarization \cite{lebeugle_very_2007} ($P = \SI{100}{\micro\coulomb\per\cm\squared}$) along the $[111]$ direction of its pseudo-cubic structure and exhibits below $T_\mathrm{N} = \SI{643}{\K}$ a G-type antiferromagnetic ordering of the iron spins with a superimposed long range cycloidal modulation in a plane containing the ferroelectric local mode vector\cite{wang_epitaxial_2003}.
In its thin film form, BiFeO$_3$ undergoes a structural transition under epitaxial strain. At low strain, it crystallizes in a so-called R-phase in the monoclinic $M_A$ structure, slightly distorted from the bulk rhombohedral structure in the R3c space group. At high strain or under certain growth conditions, BiFeO$_3$ can exhibit a structural phase with enhanced tetragonality (with a $c/a$ ratio between 1.22 and 1.25) in the monoclinic $M_C$ structure, called the super-tetragonal or T-phase in which an enhanced ferroelectric polarization has been observed and theoretically reproduced\cite{sando_multiferroic_2016}.
Many studies report the existence of intermediate mixed R-T phases in BiFeO$_3$ films, with a 3$\degree$ tilt of the monoclinic angle between the phases\cite{sando_bifeo_2014}.
Regarding functional properties, epitaxial strain has been shown to modify the ferroic properties of BiFeO$_3$ films: the ferroelectric Curie transition temperature is decreased under epitaxial strain \cite{infante_bridging_2010} while the bulk-like magnetic cycloid present at low strain is destroyed at high epitaxial strain, both tensile and compressive, leaving place to a canted antiferromagnetic state \cite{sando_crafting_2013, agbelele_strain_2016}.

Relying on the interplay between epitaxial strain and helium implantation, a full tridimensional control of strain in thin films can be achieved.
The lattice misfit between the substrate and the film's native bulk compound causes a biaxial strain in the plane of epitaxy (Fig \ref{NP-SRIM}.a and \ref{NP-SRIM}.b). Due to the in-plane clamping of the film to its substrate, helium implantation gives rise to an out-of-plane negative pressure, without modification in-plane (Fig \ref{NP-SRIM}.c).
In the case of BiFeO$_3$ thin films, helium implantation allows to trigger the transition towards the super-tetragonal phase, by enhancing tetragonality due to this out-of-plane strain tuning. It has been shown recently that large scale helium implantation on epitaxial BiFeO$_3$ films can induce the super-tetragonal phase on LSAT substrate, while enhancing tetragonality in R-like and T-like films grown on SrTiO$_3$ and LaAlO$_3$ respectively\cite{herklotz_designing_2019}. This raises important questions regarding the the transition from the R-like to T-like structure, here caused by continuously varying the helium dose: character of the transition, possibility of coexistence between the two phases, presence of interfaces, etc.
Here, by combining local and non-local probes, we bring insight on the mechanism of this R to T structural transition under helium implantation, that appears first order in character. In addition, we show that local helium implantation allows to pattern the tetragonality of the structure while preserving crystal quality, demonstrating the possibility to structurally pattern BiFeO$_3$ films.

\section{Experimental details}
%BFO samples & characterization
BiFeO$_3$ epitaxial films were grown by pulsed laser deposition on $(001)$ oriented SrTiO$_3$ (-1.5 \% lattice misfit with BiFeO$_3$) and DyScO$_3$ (-0.4 \% lattice misfit) substrates with a \SI{5}{\nm} thick SrRuO$_3$ bottom electrode as described in refs [\onlinecite{farokhipoor_conduction_2011, farokhipoor_local_2012,haykal_antiferromagnetic_2020}].
The thickness of the films was set to \SI{60}{\nm}. For the film grown on DyScO$_3$, this is thin enough to avoid strain relaxation, while for the film grown on SrTiO$_3$, the formation of in-plane ferroelastic 71$\degree$ domains induce partial strain relaxation already at \SI{60}{\nm} thickness, but their alternation allows for overall epitaxial matching with the substrate\cite{daumont_tuning_2010}.

%SRIM simultations / parameters of implantation (and images?) / amorphization discussion
Helium implantation of our films was performed using an Orion NanoFab helium ion microscope (HIM) \cite{wirtz_imaging_2019}. Various doses of 6~keV helium ions were implanted into square regions with sizes ranging from $10\times \SI{10}{\um\squared}$ to $500\times\SI{500}{\um\squared}$ (Fig. \ref{NP-SRIM}.f). We kept the doses below 1$\times$10$^{16}$ ions cm$^{-2}$, which, at the energy used, is below the threshold for structural defect formation \cite{livengood_subsurface_2009, autier-laurent_tailoring_2010, wei_hydrogen_2012}.
The implantation parameters, in particular the beam energy and the impact angle, were determined by SRIM\cite{ziegler_srim-2003_2004} (Stopping Range of Ions in Matter) and TRIM (TRansport of Ions in Matter) simulations in order to have the maximum of the implantation profile inside the film (Fig. \ref{NP-SRIM}.d,e). The samples were therefore tilted 49$\degree$ from the normal to reduce helium implantation of the bottom electrode and the substrate while also avoiding channeling through the crystal structure's easy planes. The TRIM simulations of the implanted ion distribution and depth profile $-$ taking density values of \SI{8.408}{\g\per\cm\cubed} for BiFeO$_3$, \SI{6.49}{\g\per\cm\cubed} for SrRuO$_3$ and of \SI{5.11}{\g\per\cm\cubed} for SrTiO$_3$ as taken from the literature\cite{schneeloch_neutron_2015, noauthor_srruo3_2019} $-$ are shown on Fig. \ref{NP-SRIM}.d and \ref{NP-SRIM}.e.

To probe our implanted samples, we used Raman spectroscopy, electron microscopy and X-ray diffraction (XRD) techniques.
%Raman measurements
Micro-Raman measurements were performed with 
%a \SI{633}{\nm} He-Ne laser and 
a \SI{442}{\nm} He-Cd laser in an inVia Renishaw micro-Raman spectrometer. Depth profiles were acquired by varying the focus, meaning the distance between the objective and the sample, with \SI{0.2}{\um} steps. Principle Components Analysis (PCA) and Non-negative Matrix Factorization (NMF), using the R-DATA software as described in Ref. [\onlinecite{schober_vibrational_2020}], allow to extract the layer's signal from the overall substrate's contribution.

%TEM technical details
Transmission Electron Microscopy (TEM) studies of cross-sectional samples were performed in a probe corrected FEI Titan$^{3}$ G2 60-300, working at \SI{300}{\kilo\volt} in scanning mode (STEM) using a High Angle Annular Dark-Field detector (HAADF). The microscope is equipped with a Bruker Super-X Energy-Dispersive X-ray spectroscope (EDX), used for obtaining elemental maps. Geometrical Phase Analysis (GPA)\cite{hytch_quantitative_1998} using Strain++ software, was applied on atomically resolved HAADF images in order to measure local deformations of the lattice.

%XRD & synchrotron details
The XRD synchrotron data were collected at two different synchrotron sources.
Diffracted intensity maps were recorded at the ID03 Surface Diffraction beamline of the ESRF using the 6-circle vertical diffractometer. The experiment was conducted using an incident wavelenght of 0.516 {\AA} (24~keV) and a beamsize of $43\times\SI{28}{\um\squared}$ (horizontal $\times$ vertical). For the grazing incidence experiments, an incidence angle of 3 degrees was used and allowed to produce a beam footprint on the sample surface with a size comparable to the area of the implanted region. The implanted region was aligned in the beam using a camera with a macro lense and the visible fluorescence of the sample under the x-ray beam irradiation. The data were collected using a Maxipix pixel detector and processed using the BINoculars code.
At the CRISTAL Beamline of SOLEIL Synchrotron, we performed localized XRD measurements on a 6-circle diffractometer, using a beam size of $30\times\SI{100}{\um\squared}$), with an imXpad pixel detector and a wavelength of \SI{1.2471}{\angstrom}. The implanted regions could be localized easily, due to their different 00$\emph{l}$ XRD signal, by scanning the sample spatially.
All the diffraction data are reported using the substrate lattices as reference systems.

\section{Results and discussion} 

\subsection{Raman spectroscopy}

%--------------------------------------------------------------------------------------------------
\begin{figure*}[t!]
\resizebox{\textwidth}{!}{\includegraphics{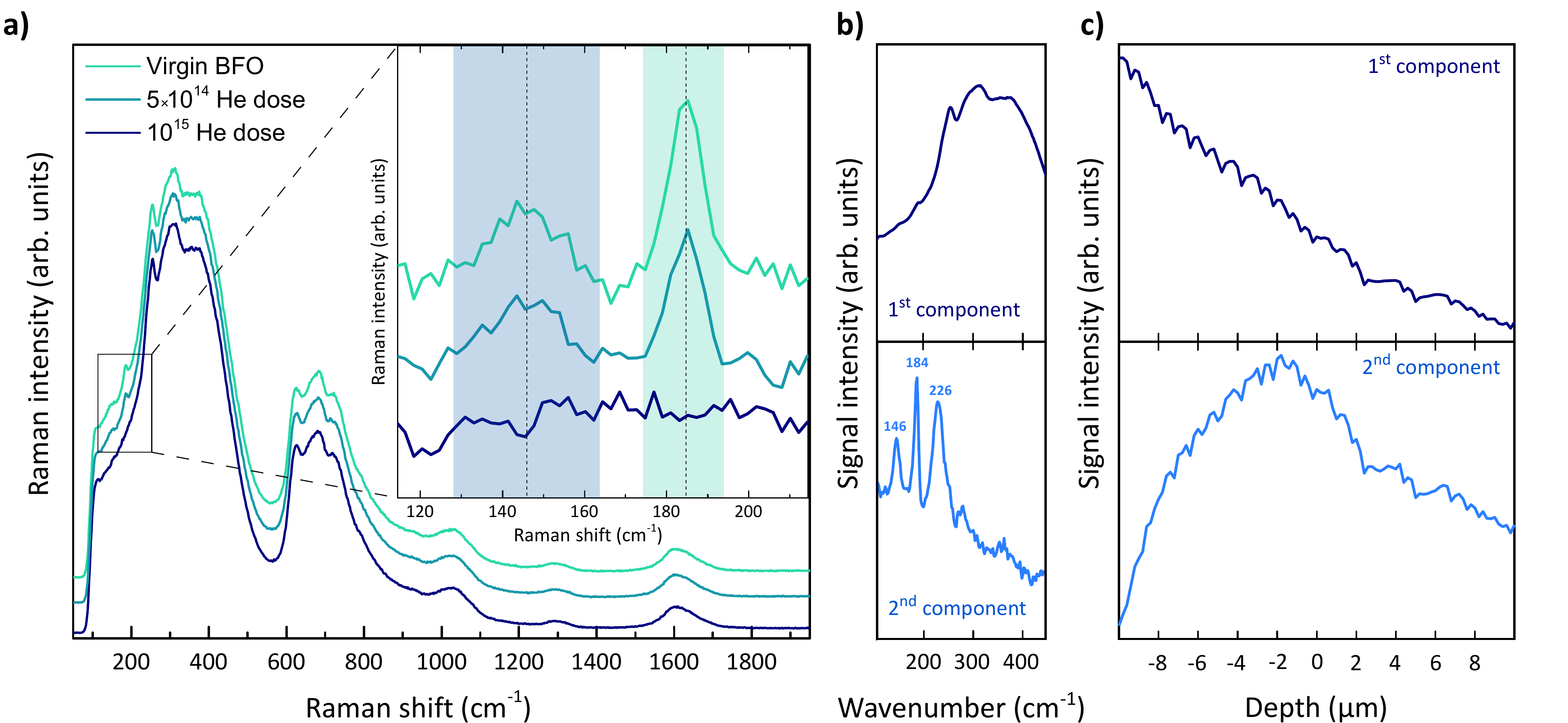}}
\caption{(a) Raman spectra performed on different regions of the BiFeO$_3$//SrTiO$_3$ film. The underlying intense signal of SrTiO$_3$ is very visible but some BiFeO$_3$ phonon modes can be singled out. The inset shows these phonon modes, with a baseline substraction. The colored regions and dotted lines are guides for the viewer.(b,c) Principal component analysis of the depth profiles on a non-implanted region of the film: we can see the component signal (b) and the associated depth profiles (c) for the first two components.}
\label{Raman_spectra}
\end{figure*}
%--------------------------------------------------------------------------------------------------

Raman spectroscopy measurements have been performed on our implanted samples to study the effect of implantation on the lattice via the phonon modes. Thin films make it naturally increasingly difficult to single out the signal of the film from the substrate contribution. However, due to the broad peaks of Raman signal of SrTiO$_3$, some sharper Raman modes from the BiFeO$_3$ layer were possible to observe in our BiFeO$_3$//SrTiO$_3$ sample. Raman spectra measured on regions with different doses are shown in Fig. \ref{Raman_spectra}.a. Two low energy phonon modes can be seen emerging from the substrate's signal at 146 and 184 cm$^{-1}$. These phonon modes correspond to Raman excitations with $A_1(LO)$ symmetry, usually measured around 145-147 and 176-180 cm$^{-1}$ respectively\cite{fukumura_raman_2007, cazayous_electric_2007}.

Furthermore, we performed Raman depth profiles by continuously varying the distance between the sample and the Raman objective, following the approach described in Ref. [\onlinecite{schober_vibrational_2020}]. Fig. \ref{Raman_spectra}.b and Fig. \ref{Raman_spectra}.c, present the Principal Component Analysis (PCA) of the Raman signal obtained on the non-implanted part of the film. Fig. \ref{Raman_spectra}.b shows the signal associated to the first two components while Fig.\ref{Raman_spectra}.c shows their intensity profile as a function of the depth.
The first component of the PCA visibly corresponds to the SrTiO$_3$ signal: it has the same shape than the well-known SrTiO$_3$ 2$^{nd}$-order Raman signal (Fig. \ref{Raman_spectra}.b, top) while its intensity continuously increases when the distance between the sample and the objective decreases (Fig \ref{Raman_spectra}.c, top). This is in agreement with the signal coming from the substrate, further away from the objective.
The second component of the PCA presents a signal where 3 narrow peaks can be observed at 146, 184 and 226~cm$^{-1}$ (Fig. \ref{Raman_spectra}.b, bottom). These peaks corresponds to the two BiFeO$_3$ Raman modes observed in the spectra of Fig. \ref{Raman_spectra}.a, with an additional peak at 226~cm$^{-1}$ also associated to an $A_1(LO)$ Raman mode of BiFeO$_3$ in the literature\cite{fukumura_raman_2007,cazayous_electric_2007}. The intensity depth profile of this second component presents a maximum in the depth range corresponding to the film (Fig. \ref{Raman_spectra}.c, bottom) which confirms that we can indeed attribute it to the Raman signal of the BiFeO$_3$ film.

%Principal component analysis of depth profiles, performed by continuously varying the focus distance of our Raman set-up, confirm that those two peaks indeed correspond to the film. We can see (Fig. \ref{Raman_spectra}.b,c) that the component of the Raman signal featuring these two peaks is present in a depth range corresponding to the BiFeO$_3$ film.

By comparing the evolution of the Raman signal with the implanted helium dose, we observe  that the Raman signal of the BiFeO$_3$ film  disappears with increasing dose while not shifting in frequency. Indeed, the two phonon modes are visible only for the virgin film and the lowest dose (5$\times$10$^{14}$ He cm$^{-2}$, Fig. \ref{Raman_spectra}.a). The Raman spectrum of BiFeO$_3$ could not be seen for the films deposited on DyScO$_3$ substrate, due to a stronger substrate contribution. 

Several causes can be envisioned to explain the disappearance of the Raman signal. In the context of BiFeO$_3$ in particular, the Raman signal of the super-tetragonal phase is weaker by one order of magnitude than the signal for the rhombohedral-like structure \cite{iliev_polarized_2010}. The disappearance of the Raman signal under implantation can therefore be a signature of a transition towards an increased tetragonality, as we would expect (cf. Fig. \ref{NP-SRIM}.c and Ref. [\onlinecite{herklotz_designing_2019}]). Alternatively, a vanishing Raman spectrum can be associated to a transition towards a metallic character, which in BiFeO$_3$ is known to happen at very high temperatures and pressure. In this particular context, it is also conceivable that an increase in band gap $-$ i.e. a decrease in absorption $-$ would lead to a decrease in intensity of the Raman spectrum as compared to the substrate. Finally, we cannot exclude an effect directly due to the presence of He in the BiFeO$_3$ lattice, decreasing the Raman intensity.
%Here, we note that the vanishing of the Raman spectrum does not correspond to the transition to the supertetragonal phase identified by XRD but occurs at much lower doses.

%\subsection{Reversibility of helium implantation}

\subsection{Transmission Electron Microscopy}

%--------------------------------------------------------------------------------------------------
\begin{figure*}[htpb]
\resizebox{\textwidth}{!}{\includegraphics{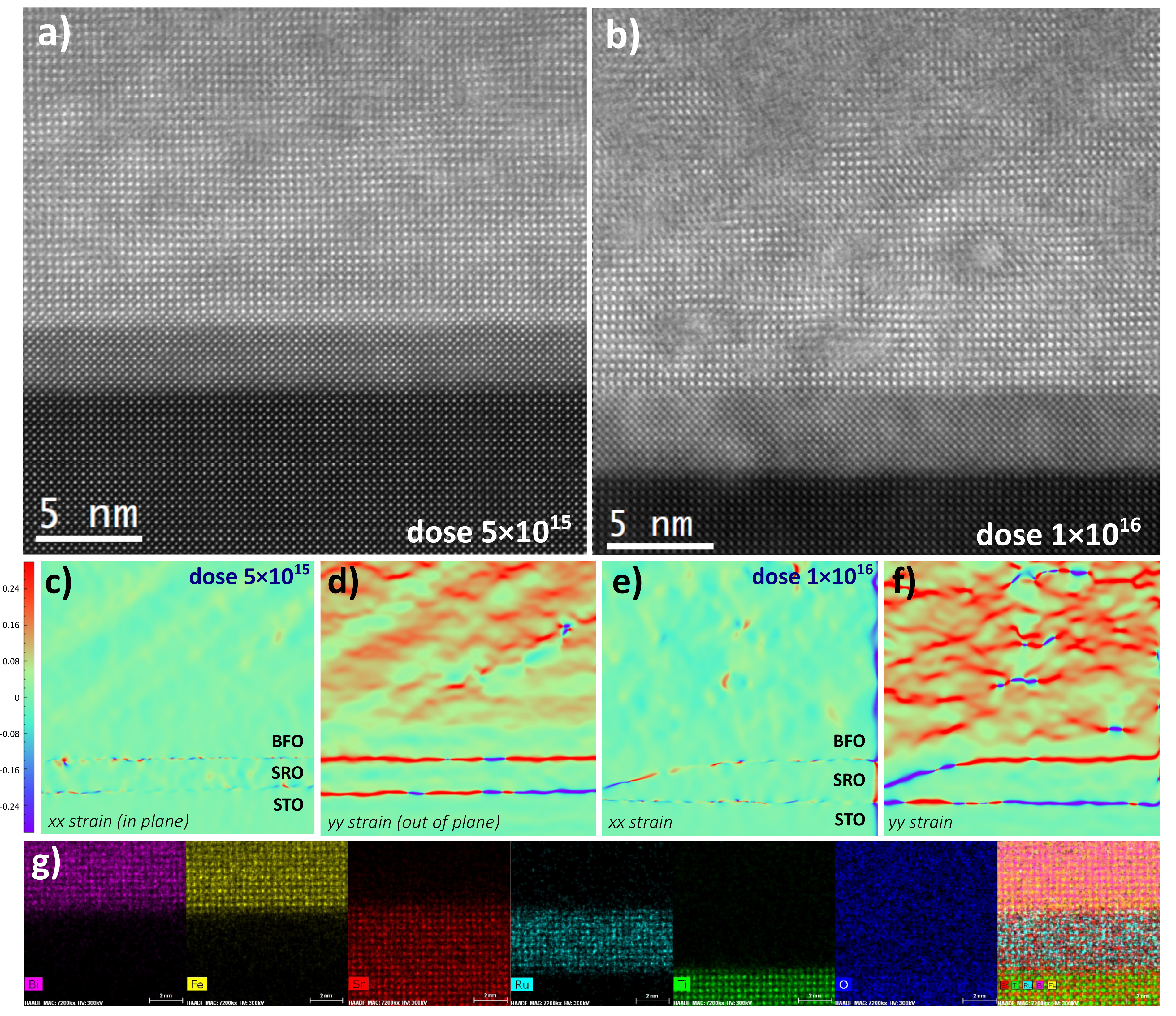}}
\caption{(a,b) STEM images (HAADF) of BiFeO$_3$/SRO//SrTiO$_3$ samples implanted with doses 5$\times$10$^{15}$ (a) and 1$\times$10$^{16}$ (b) He cm$^{-2}$. (c-f) GPA (Geometrical Phase Analysis) of the strain in the epitaxial plane (c,e) and out of plane (d,f), using the SrTiO$_3$ substrate as a reference, of the samples with dose 5$\times$10$^{15}$ (c,d) and 1$\times$10$^{16}$ (e,f) He cm$^{-2}$.(g) EDX chemical analysis of the interface in the sample implanted with 5$\times$10$^{15}$ He cm$^{-2}$. }
\label{TEM}
\end{figure*}
%--------------------------------------------------------------------------------------------------

To gain more insight on the effect of implantation at the local scale, Transmission Electron Microscopy was performed on prepared samples from a BiFeO$_3$ layer grown on a SrTiO$_3$ (001) substrate, with doses 5$\times$10$^{15}$ and 1$\times$10$^{16}$ He cm$^{-2}$.

The High-Angle Annular Dark Field (HAADF) STEM images of the BiFeO$_3$/SrTiO$_3$ interfaces, in the two implanted regions are shown on Fig. \ref{TEM}.a and \ref{TEM}.b. No amorphization of the BiFeO$_3$ layer is visible under implantation for the region with 5$\times$10$^{15}$ He cm$^{-2}$ dose : the epitaxial columns are clearly visible throughout the film although some structural defects are present (as expected under implantation).
%Moreover, the 45$\degree$ domain walls as described in [\onlinecite{daumont_tuning_2010}] and observed in [\onlinecite{jang_domain_2009}] seem also visible (Fig. \ref{TEM}.a).
When the implanted dose is increased to a higher level, however, some regions of amorphization are observed (Fig.~\ref{TEM}.b). Their characteristic size is of the order of the nanometer, and is reminiscent of the observation of defect ``nano-bubbles" previously reported upon He implantation on silicon \cite{livengood_subsurface_2009}.

To gain insight on the strain induced locally by helium implantation, Geometric Phase Analysis (GPA) has been performed on samples prepared with 5$\times$10$^{15}$ and 1$\times$10$^{16}$ He cm$^{-2}$ doses, taking as a reference the SrTiO$_3$ lattice to extract local deformations. We see in Fig.~\ref{TEM}.c and \ref{TEM}.e the in-plane strain for both doses, while the out-of-plane strain is shown on Fig.~\ref{TEM}.d and \ref{TEM}.f. While it is apparent from the GPA strain maps that local deformations in-plane are very small, we observe stronger out-of-plane deformations, consistent with an out-of-plane swelling induced by implantation. The strain seems to increase with the dose between Fig.~\ref{TEM}.d and \ref{TEM}.f, as expected for an enhanced tetragonality.

Furthermore, we extracted the average lattice parameters by performing Fourier transform of the STEM images in the two regions with 5$\times$10$^{15}$ and 1$\times$10$^{16}$ He cm$^{-2}$ doses as well as in a pristine -non-implanted- region as a reference. They are given in Table \ref{lattice_TEM}. Each lattice parameter was estimated from 10 images and the error bars arise from the standard deviation of the measurements. We can clearly see that the in-plane lattice parameter remains unchanged while a c-lattice expansion, increasing with the helium dose, is observed with implantation. Moreover, we can observe that in the region implanted with the highest dose, the c-axis lattice parameter extracted by Fourier transform has a much broader error bar, consistent with a loss of crystallinity of the film as visible from Fig. \ref{TEM}.b.

\begin{table}[t!]
    \centering
    \begin{tabular}{l|c|c|c}
        \hline \hline
        Sample & a & c & c/a$_{\small{STO}}$\\
        \hline
        STO substrate & 3.79 $\pm$ \SI{0.02}{\angstrom} & 3.92 $\pm$ \SI{0.03}{\angstrom} & \\
        BFO pristine & 3.82 $\pm$ \SI{0.02}{\angstrom} & 4.02 $\pm$ \SI{0.02}{\angstrom} & 1.03\\
        BFO (5$\times$10$^{15}$ He cm$^{-2}$) & 3.79 $\pm$ \SI{0.01}{\angstrom} & 4.38 $\pm$ \SI{0.05}{\angstrom} & 1.12\\
        BFO (1$\times$10$^{16}$ He cm$^{-2}$) & 3.78 $\pm$ \SI{0.04}{\angstrom} & 4.57 $\pm$ \SI{0.10}{\angstrom} & 1.17\\
        \hline \hline  
    \end{tabular}
    \caption{Lattice parameters extracted from the STEM measurements. The discrepancy between the usual SrTiO$_3$ cubic lattice parameter of \SI{3.905}{\angstrom} and the extracted ones in the a direction comes from the calibration of the microscope combined with sample drift. We chose to keep the values as extracted from the images, the values in the c-axis being in agreement with literature. The last column shows the c/a ratio, calculated with the a lattice parameter of SrTiO$_3$ from literature: a$_{STO}$=\SI{3.905}{\angstrom}}
    \label{lattice_TEM}
\end{table}

%--------------------------------------------------------------------------------------------------
\begin{figure*}[t!]
\resizebox{\textwidth}{!}{\includegraphics{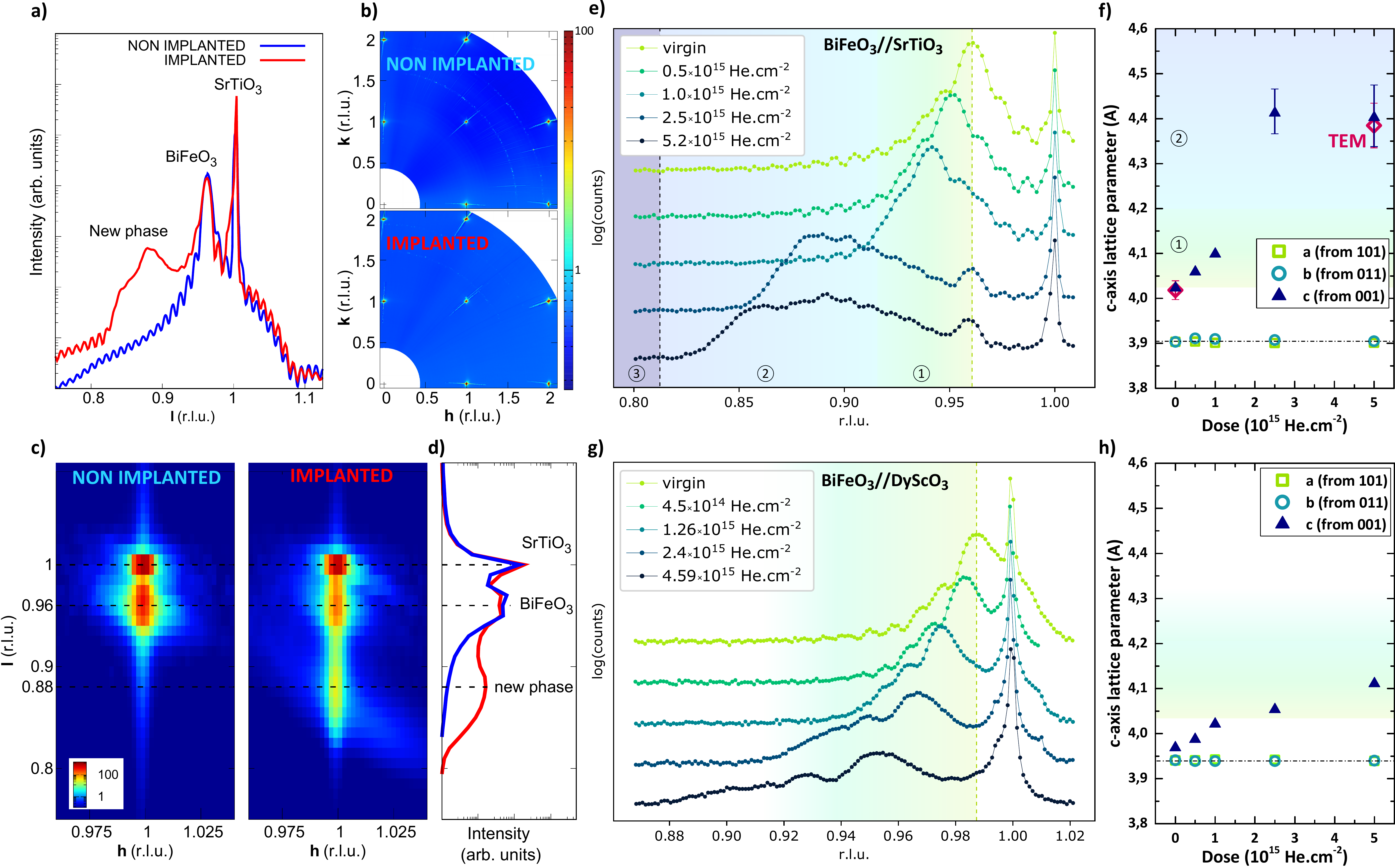}}
\caption{X-ray diffraction results of non-implanted and implanted (with a 5$\times$10$^{15}$ He cm$^{-2}$ dose) regions of a BiFeO$_3$//SrTiO$_3$ thin film. (a) Specular reflectivity. (b) (\emph{h},\emph{k}) reciprocal space maps integrated in the -0.05$<$\emph{l}$<$0.05 range.
(c) (\emph{h},\emph{l}) reciprocal space maps measured at \emph{k} = 0.96 and (d) the corresponding 1-D plot by integrating the intensity of the map contained in the 0.98$<$\emph{h}$<$1.02 range. (e) $\theta(2\theta)$ scans of regions implanted with various helium doses. (f) Extracted out-of-plane (blue triangles) and in-plane (green circles) lattice parameters (from 001, 101 and 011 Bragg peaks) as a function of the helium dose. The out-of-plane lattice parameters estimated from STEM observations are added as red diamonds. Similar experiments for a BiFeO$_3$//DyScO$_3$ sample with (g) $\theta(2\theta)$ scans of regions implanted with various helium doses and (h) the extracted out-of-plane (blue triangles) and in-plane (green circles) lattice parameters (from 001, 101 and 011 Bragg peaks) as a function of the helium dose.}
\label{XRD_tot}
\end{figure*}
%--------------------------------------------------------------------------------------------------

Finally, Energy Dispersive X-Ray Spectroscopy (EDX) chemical analysis of the atoms at the interface between the substrate and the film (with a \SI{5}{\nm} thick SrRuO$_3$ bottom electrode) has been performed and is presented in Fig.~\ref{TEM}.g. The chemical compositions of the substrate, the bottom electrode and the layer are well resolved and the compositional interfaces correspond to those of the HAADF image with no indication of chemical intermixing. Furthermore, no migration of the atoms of the layer to the bottom electrode or substrate is observed under implantation.

\subsection{Synchrotron XRD}

To determine the effect of helium implantation on the structure of our BiFeO$_3$ layers, we performed micro-XRD, using specular reflectivity and grazing incidence geometry.
Fig. \ref{XRD_tot}.a shows the specular reflectivity curve measured inside an implanted region (with 5$\times$10$^{15}$ He cm$^{-2}$ dose) on a BiFeO$_3$ film grown on SrTiO$_3$ substrate. It shows, in addition to the (001) Bragg peaks of the SrTiO$_3$ substrate and the BiFeO$_3$ layer, a new phase with a $c$-lattice expansion, revealed by the new peak appearing at \emph{l}=0.89 r.l.u (Fig. \ref{XRD_tot}.a).

From the in-plane diffracted intensity maps measured at different \emph{l} values it was possible to verify that all the phases have the same in-plane lattice parameters of the substrate, as observed locally from the TEM results. A comparison between the \emph{hk} maps collected at \emph{l}=0.96 for the implanted and non-implanted regions (Fig. \ref{XRD_tot}.b) shows no difference in the positions of the BiFeO$_3$ Bragg peaks, indicating no modification of the in-plane lattice parameter as a consequence of the implantation process: as expected, the clamping to the substrate locks the film parameters in the epitaxial plane.
Fig. \ref{XRD_tot}.c shows the \emph{hl} maps around the (001) Bragg peak of the SrTiO$_3$ substrate outside and inside an implanted region (with 5$\times$10$^{15}$ He cm$^{-2}$ dose). The $(001)$ peak of the substrate and, at lower \emph{l}, of the BiFeO$_3$ layer, are clearly visible while a new phase, with a $c$-lattice expansion of 7.9\% (from 4.072 to \SI{4.395}{\angstrom}) is observed in the implanted region, consistent with the values extracted from the TEM images. Fig. \ref{XRD_tot}.d shows a close-up of the (10\emph{l}) rod around the (001) region, in which the coexistence of the new structural phase and the pristine BiFeO$_3$ layer are visible.
This can be explained by a mixing of the two structural phases, the pristine R-like phase and the new phase with enhanced tetragonality in the implanted region.
Alternatively, the R-type signal could also arise from neighboring non-implanted regions, partly probed due to the X-ray spot size that was slightly larger than the implanted region.

In order to study the conditions of appearance of this new phase under helium implantation, we performed implantations with increasing doses. The $\theta(2\theta)$ XRD scans, around the $(001)$ Bragg peak of the SrTiO$_3$ substrate for doses ranging between 5$\times$10$^{14}$ and 5$\times$10$^{15}$ He cm$^{-2}$ are shown on Fig.~\ref{XRD_tot}.e. We observe the $(001)$ peak of the BiFeO$_3$ layer progressively shifting towards higher $c$-axis lattice parameters with dose. Under 2.5$\times$10$^{15}$ He cm$^{-2}$ (region \raisebox{.9pt}{\textcircled{\raisebox{-.9pt}{1}}}, in green Fig. \ref{XRD_tot}.e), the layer's (001) peak shifts continuously. At 2.5$\times$10$^{15}$ He cm$^{-2}$, however, we observe an abrupt change in the (001) Bragg signal of the layer (region \raisebox{.9pt}{\textcircled{\raisebox{-.9pt}{2}}}, in blue Fig.~\ref{XRD_tot}.e), with a strong shift of the layer peak, that we interpret as the appearance of the new phase previously observed in Fig.~\ref{XRD_tot}.c,d and the onset of the R-like to T-like BiFeO$_3$ structural transition.
In this new phase (at 2.5$\times$10$^{15}$ He cm$^{-2}$ dose and above), we also observe the presence of a peak at the position of the $(001)$ peak of the virgin (non-implanted) BiFeO$_3$ layer, as observed in Fig. \ref{XRD_tot}.c,d. In this measurement, the spot size was well below the size of the implanted regions and its centering by means of (x,y) scans ruled out the possibility of signal coming from non-implanted BiFeO$_3$.
The presence of the $(001)$ pristine-like BiFeO$_3$ peak therefore unambiguously points to a phase mixture in the implanted region. 

The lattice parameters were extracted directly from the $(001)$, $(101)$ and $(011)$ XRD peaks of the film in the different regions. They are shown, as a function of the implanted helium dose, in Fig.~\ref{XRD_tot}.f, the dotted line marking the lattice parameter of the SrTiO$_3$ cubic substrate. As observed previously, the in-plane lattice parameters are not modified under implantation. 

To investigate the combined effect of epitaxial strain and implantation, we performed implantations with the same helium doses on a film synthesized on a DyScO$_3$ susbtrate, which has a lattice parameter closer to BiFeO$_3$ (-0.4 \% lattice misfit\cite{haykal_antiferromagnetic_2020}). Fig.~\ref{XRD_tot}.g shows the $\theta(2\theta)$ XRD scans, around the $(001)$ Bragg peak of the DyScO$_3$ substrate for doses ranging between 5$\times$10$^{14}$ and 5$\times$10$^{15}$ He cm$^{-2}$. We see the $(001)$ peak of the BiFeO$_3$ layer progressively shifting with dose, showing an increase of the out-of-plane lattice parameter.
Contrary to the results on the SrTiO$_3$ substrate, no abrupt modification is observed and the $c$ lattice parameter (reported in Fig.~\ref{XRD_tot}.h) is in the same range than in the \raisebox{.9pt}{\textcircled{\raisebox{-.9pt}{1}}} region of the Fig. \ref{XRD_tot}.f on the SrTiO$_3$ sample. This indicates a strained R-like structure with no phase transition.
As on SrTiO$_3$,  we observe no modification of the in-plane $a$ and $b$ lattice parameters, locked by the DyScO$_3$ lattice parameter (dotted line, Fig. \ref{XRD_tot}.h).

\subsection{Discussion}

By combining Raman spectroscopy, electron microscopy and X-ray diffraction, we could access the structural changes due to implantation at different length scales, allowing for a more complete insight into the transition between the rhombohedral and super-tetragonal phases of BiFeO$_3$ films than previously reported. Specifically, helium implantation is a way to observe the transition with a continuously varying parameter, as opposed to epitaxial strain that typically takes only a few discrete values fixed by the substrates.

Our TEM measurements, in particular, clearly show both the elongation of the unit cell and the partial amorphization. We indeed observe amorphized regions at 1$\times$10$^{16}$ He cm$^{-2}$ dose, while at the same time, the c-axis lattice parameter extracted by Fourier transform from the non-amorphized regions present a ratio with the a-axis lattice parameter which is close to the $c/a$ ratio of the super-tetragonal phase\cite{sando_multiferroic_2016}. It therefore seems that despite the structural damage due to the loss of crystallinity of some regions, we continue to enhance the tetragonality of the still crystalline regions of the film, towards the T-phase. It is not possible to know if amorphization was also present in the previous study on He-implantation of BiFeO$_3$ films using a large scale ion implantor\cite{herklotz_designing_2019} as XRD is only sensitive to the crystalline regions.
%It is very well possible that such amorphization also occurred in previous studies [\onlinecite{herklotz_designing_2019}] but remains unnoticed, as the XRD is sensitive only to the crystalline regions. 
In our XRD data, where the dose was kept low enough to avoid amorphization, the elongation of the unit cell does not reach the $c/a$ typical of the super-tetragonal phase. Whether or not it is possible to fully transform into the super-tetragonal phase on SrTiO$_3$ while remaining fully crystalline, is still to be investigated. 

The picture provided by our TEM and XRD data shows significant differences when compared to the same phase transition reported in Ref.~\onlinecite{herklotz_designing_2019} on (LaAlO$_3$)$_{0.3}$(Sr$_2$TaAlO$_6$)$_{0.7}$  (LSAT) and SrTiO$_3$. We observe the transition towards a mixed phase with an enhanced tetragonality both at lower epitaxial strain and at lower doses: Herklotz et al.\cite{herklotz_designing_2019} observe a structural transition towards the T phase on LSAT, with a -2.6\% lattice misfit and a 6$\times$10$^{15}$ He cm$^{-2}$ dose, whereas we already see a similar behaviour with enhanced tetragonality on SrTiO$_3$ where the lattice misfit is only -1.5\% at a 2.5$\times$10$^{15}$ He cm$^{-2}$ dose. These discrepancies are not negligible, and their origin is for the moment unclear. They could be due to an inaccurate estimation of the implanted dose or to a different local behaviour, resulting in a strain gradient, that may be due to the difference in implantation technique. In particular, the presence of interfaces between implanted and non-implanted regions, specific to the patterning possibilities of our implantation technique, can impact the strain relaxation mechanisms and account for a higher applied strain with respect to the dose as compared to large scale implantation techniques. Moreover, the thickness and possible domain structure differences can give rise to different stress field landscapes, that may account for the differences in our study. 
%Indeed, the strain effect due to implantation seems to be domain-dependent, as seen from the GPA results of Fig. \ref{TEM}.(c,d). 
Furthermore, the energy of the accelerated He ions with the He-FIB we use is higher than with a large scale ion implanter as used in Ref.~[\onlinecite{herklotz_designing_2019}]. Special attention should be paid to these aspects in future studies.

A noticeable difference also lies in the way the transition proceeds with increasing dose. In Ref. [\onlinecite{herklotz_designing_2019}], the transition is described as appearing continuously under implantation with a progressive shift of the Bragg peak (cf. Fig. 1.b of the Ref. [\onlinecite{herklotz_designing_2019}] for the film grown on LSAT). They further support the scenario of a continuously rotating polarization based on PFM and SHG observations. In our measurements, in contrast, the emergence of the T-like phase appear the be step-like with a visible phase coexistence, reminiscent of a first-order transition. We believe this is not, in fact, incompatible with their raw XRD data, where a splitting of the Bragg peak can be seen at intermediate doses. We hypothesize that the continuous character observed in PFM and SHG could result from an averaging effect that does not reflect the details of the local picture.

Finally, the comparison between Raman and XRD data sheds light on the disappearance of the Raman spectrum. We can see that the Raman signal disappears at a dose (1$\times$10$^{15}$ He cm$^{-1}$) where the BiFeO$_3$ is still clearly in a slightly elongated R phase. Indeed, the $\theta(2\theta)$ XRD scan (Fig. \ref{XRD_tot}.e) still shows a narrow 001 Bragg peak, shifted from the pristine BiFeO$_3$, but still showing no sign of transition from the rhombohedral-like structure. Therefore, the vanishing of the Raman signal cannot be attributed to the lower Raman intensity known for the T phase, and cannot be understood as a signature of the transition towards the tetragonal-like structure. Instead, we suggest that it reflects a decrease of Raman susceptibility and polarizability due to the insertion of helium ions. This might be significant also for the dielectric properties of the implanted films.

\section{Conclusion}

Helium implantation has arisen recently as a powerful technique with the potential to modify and tune the strain state in a perovskite film. We have shown here, in particular, that local helium implantation by means of a helium ion microscope can enhance tetragonality in (001)-oriented BiFeO$_3$ thin films, increasing the out-of-plane lattice parameter up to 4\% in films grown on DyScO$_3$ and up to 9\% in films grown on SrTiO$_3$, inducing locally a structural phase transition towards a mixed R-T phase.

Combined Raman spectroscopy, electron microscopy and X-Ray diffraction measurements allowed us to probe the structural changes both at the large scale and locally.
Our local (TEM) study allowed us to determine the threshold dose for which amorphization starts to appear in implanted regions, while observing elongations of the unit cell along the c-axis up to the values nearing the super-tetragonal lattice distortion. 
%We could determine the dose range necessary to preserve a good crystallinity while still inducing these important lattice constant changes.
Synchrotron XRD performed in the locally implanted regions allowed us to probe the onset of the transition between the rhombohedral-like structure and the super-tetragonal phase. Our data suggest that this transition appears as a first-order transition, with an abrupt jump of the c-axis expansion at a threshold dose (2.5$\times$10$^{15}$ He cm$^{-1}$ on SrTiO$_3$), which opens the discussion for understanding the underlying mechanisms at stake in this structural transition under implantation.

More generally, this work demonstrates the use of a helium microscope as a powerful mean for strain-engineering by local helium implantation and opens technical possibilities for property tuning and patterning.

\section*{Acknowledgments}

This work was supported by the Fond National de la Recherche (FNR) of Luxembourg (PEARL CO-FERMAT, Project code FNR/P12/4853155/Kreisel). We acknowledge SOLEIL for provision of synchrotron radiation at CRISTAL beamline (proposal number 20190831). TEM sample preparation and observation was carried out within the MATMECA consortium and supported by the ANR under contract number ANR-10-EQPX-37. It has benefited from the facilities of the Laboratory MSSMat, CNRS, CentraleSupélec, Université Paris-Saclay, France. The author Inma Peral is supported by the National Research Fund of Luxembourg (Grant No FNR-Inter2015/LRSF). Saeedeh Farokhipoor acknowledges ﬁnancial support from  a VENI grant (016.veni.179.053) of the Netherlands Organisation for Scientiﬁc Research (NWO).

\bibliography{NPBFObib}

\end{document}